\documentclass[a4paper,12pt]{article}
\pdfoutput=1

\usepackage{graphicx}
\usepackage{mathtools}
\usepackage{amssymb}
\usepackage{amsfonts}
\usepackage[footnotesize]{caption}
\usepackage[font=scriptsize]{subcaption}
\usepackage{color}
\usepackage{braket}
\usepackage{cite}
\usepackage{hyperref}
\usepackage{url}
\usepackage{multirow}
\usepackage{relsize}
\usepackage{fullpage}
\usepackage{makecell}
\usepackage{blkarray}
\usepackage{fullpage}

\usepackage{tikz-feynman}
\tikzfeynmanset{compat=1.0.0}

\newcommand{\newc}{\newcommand}
\newc{\be}{\begin{equation}}
\newc{\ee}{\end{equation}}
\newc{\bea}{\begin{eqnarray}}
\newc{\eea}{\end{eqnarray}}
\newc{\simlt}{~\mbox{\smaller\(\lesssim\)}~}
\newc{\simgt}{~\mbox{\smaller\(\gtrsim\)}~}

\newcommand{\pmatr}[1]{\begin{pmatrix} #1 \end{pmatrix}}

\begin{document}

\begin{titlepage}
\begin{center}
{\bf\Large
\boldmath{
Origin of Yukawa couplings for Higgs and leptoquarks 
}
} \\[12mm]
Ivo~de~Medeiros~Varzielas$^{\dagger}$
\footnote{E-mail: \texttt{ivo.de@udo.edu}},
Stephen~F.~King$^{\star}$%
\footnote{E-mail: \texttt{king@soton.ac.uk}},
\\[-2mm]
\end{center}
\vspace*{0.50cm}
\centerline{$^{\dagger}$ \it
CFTP, Departamento de F\'{i}sica, Instituto Superior T\'{e}cnico, Universidade de Lisboa,}
\centerline{ \it
 Avenida Rovisco Pais 1, 1049 Lisboa, Portugal}
\centerline{$^{\star}$ \it
School of Physics and Astronomy, University of Southampton,}
\centerline{\it
SO17 1BJ Southampton, United Kingdom }

\vspace*{1.20cm}

\begin{abstract}
{\noindent
We propose a model in which the Yukawa couplings of Higgs doublets are related to
the couplings of the chiral fermions to a scalar leptoquark triplet.
This is due to their common origin
via mixing with a vector-like family distinguished by a discrete $Z_5$ symmetry, under which only the three chiral families
are neutral. The model predicts lepton non-universality in $B$ to $K$ decays, depending on the leptoquark mass,
$V_{ts}$ and $m_\mu/m_\tau$. The model can only consistently explain the anomalies in $R_{K^{(*)}}$ for a
leptoquark mass close to the collider lower bound of about $1~\text{TeV}$. 
Constraints from $B_s - \bar{B}_s$ mixing and eventually $\tau \to \mu \gamma$ become relevant
for low leptoquark masses and large couplings, while $\mu \to e \gamma$ remains automatically under control due to the absence of leptoquark couplings to the electron in this model.
}
\end{abstract}
\end{titlepage}

\section{Introduction}

In the Standard Model (SM), the charged fermion masses and the Cabibbo-Kobayashi-Maskawa (CKM) quark mixing matrix,
with entries
$V_{td}$, $V_{ts}$, $V_{tb}$, etc.,
arise from Yukawa couplings to a Higgs doublet, while the 
origin of the neutrino masses and the Pontecorvo-Maki-Nakagawa-Sakata (PMNS) is unknown.
However, even the charged fermion sector is unsatisfactory since the Yukawa couplings are essentially free parameters and provide no insight into the flavour puzzle. Many theories of flavour Beyond the Standard Model (BSM) try to explain the
Yukawa couplings as arising from non-renormalisable operators suppressed by some heavy mass scale(s), but the magnitude
of such flavour scale(s) is unknown and can vary from the Planck scale to the electroweak scale. If such flavour scale(s) are close to the electroweak scale, then one may hope to see some hint of the new physics in flavour violating observables.

One example of a flavour violating observable is the recent 
indication for semi-leptonic $B$ decays deviating from $\mu - e$ universality differing from the SM prediction
\cite{Descotes-Genon:2013wba,Altmannshofer:2013foa,Ghosh:2014awa}.
LHCb Collaboration along with other experiments observe deviations from the SM in decays $B\rightarrow K^{(*)}l^+l^-$, as seen in the ratios of $\mu^+ \mu^-$ to $e^+ e^-$ final states $R_K$ \cite{Aaij:2014ora} and $R_{K^*}$ \cite{Aaij:2017vbb}, 
at $\sim70\%$ of SM values, consisting in deviation by $4\sigma$. Additionally there is also the observable $P'_5$ angular dependence,
and the $B\rightarrow \phi \mu^+ \mu^-$ mass distribution in $m_{\mu^+ \mu^-}$. 

After $R_{K^*}$ was measured \cite{Aaij:2017vbb}, phenomenological analyses prefer
explanations with an operator
$\bar b_L\gamma^{\mu} s_L \, \bar \mu_L \gamma_{\mu} \mu_L$,
an operator $\bar b_L\gamma^{\mu} s_L \, \bar \mu \gamma_{\mu} \mu$,
or some linear combination of both operators,
in each case with a dimensionful coefficient $\Lambda^{-2}$ where $\Lambda \sim 31.5$ TeV (see e.g. \cite{Glashow:2014iga, Calibbi:2015kma, Descotes-Genon:2015uva, Capdevila:2017bsm, DAmico:2017mtc, Hiller:2017bzc, Geng:2017svp, Ciuchini:2017mik, Ghosh:2017ber, Alok:2017jaf, Alok:2017sui, Bardhan:2017xcc}).
The operator $\bar b_L\gamma^{\mu} s_L \, \bar \mu_L \gamma_{\mu} \mu_L$ can arise from $S_3$, an $SU(2)_L$ triplet scalar leptoquark \cite{Hiller:2017bzc}:
\be
\lambda^{ij}S_3Q_iL_j\equiv \lambda^{ij}S_3^{\beta \gamma}Q^{\alpha}_i(i\sigma_2)^{\alpha \beta} L_j^{\gamma} \,,
\label{S3}
\ee
where we show $\alpha, \beta, \gamma $ ($SU(2)_L$ indices) only
on the right-hand side, and where the chiral families SM fermions $SU(2)_L$ doublets in two component Weyl notation are denoted as $Q_i$ and $L_j$ (with $i,j=1,2,3$).
$\bar b_L\gamma^{\mu} s_L \, \bar \mu_L \gamma_{\mu} \mu_L$ 
(and other operators) then appears at tree-level from $S_3$ (with a Fierz transformation).
However, the introduction of such a leptoquark only deepens the mystery of the flavour problem in the SM, by introducing
yet more undetermined Yukawa couplings, this time to the leptoquark. It would clearly be nice to be able to link
such leptoquark Yukawa couplings somehow to the usual Yukawa couplings to the Higgs doublet, in order to make such theories more predictive.

One attractive scenario is that the usual Yukawa couplings of the SM, as well as the new Yukawa couplings to the leptoquark,
could have a common origin, namely due to operators which are mediated by a fourth vector-like family with TeV scale masses \cite{Ferretti:2006df, Calibbi:2008yj}.
This has already been proposed in the framework of $Z'$ models, where only the fourth family (not the three chiral families)
can carry a gauged $U(1)'$. Due to mixing with the fourth family, the three chiral families develop couplings to the massive $Z'$ gauge boson with effective non-universal couplings, which can account for  $R_{K^*}$ \cite{King:2017anf}. 
This idea has been further explored in F-theory models with non-universal gauginos \cite{Romao:2017qnu};
$SO(10)$ models (addressing also the issue of neutrino mass) \cite{Antusch:2017tud};
$SU(5)$ models (with a focus on the Yukawa relation $Y_e\neq Y_d^T$)  \cite{CarcamoHernandez:2018aon};
and $Z'$ portal models with a coupling to a fourth-family singlet Dirac neutrino Dark Matter,
discussing all phenomenological constraints \cite{Falkowski:2018dsl}. A similar idea was also considered in \cite{Raby:2017igl}, where
phenomenological implications such as 
the muon $g-2$ and $\tau \rightarrow \mu \gamma$ were also considered.
In general, the literature proposing explanations for $R_{K^{(*)}}$ is huge, but relatively few papers are concerned with its possible connection with flavour. The connection with Yukawa couplings has been considered in \cite{Varzielas:2015iva, Pas:2015hca, Hiller:2017bzc, Guo:2017gxp, Aloni:2017ixa,King:2018fcg, Hati:2018fzc, deMedeirosVarzielas:2018bcy, Grinstein:2018fgb, deMedeirosVarzielas:2019lgb}.
In this context, a connection between $R_{K^{(*)}}$ and the origin of fermion Yukawa couplings was recently studied in a $Z'$ model \cite{King:2018fcg}.

Recently we considered a model with the scalar leptoquark $S_3$, in which 
the physics that generates the Yukawa couplings is related to the structure of the couplings that accounts for $R_{K^{(*)}}$
\cite{deMedeirosVarzielas:2018bcy}.
The model considered in 
\cite{deMedeirosVarzielas:2018bcy} is based on
having a fourth vector-like family distinguished by a discrete $Z_2$ and with the additional scalar leptoquark
$S_3$ (and $SU(2)_L$ triplet) being odd under this $Z_2$ parity. This combinations leads to a model where the explanation for $R_{K^{(*)}}$ is linked with the origin of the Yukawa couplings: the leptoquark couplings to SM fermions (Eq.\ref{S3}) are in this case mediated by the fourth family vector-like fermions, and are related to CKM entries and mass ratios of SM fermions. In this model \cite{deMedeirosVarzielas:2018bcy}, the leptoquark Yukawa couplings arise at the same order as the Yukawa couplings, suppressed by only one power of the vector-like family mass, leading to relatively heavy leptoquark masses above the $O(1)~\text{TeV}$ scale, while facing a severe challenge from $\mu \to e \gamma$.

In the present paper we consider a similar 
framework to what was proposed in \cite{deMedeirosVarzielas:2018bcy}, but suppose that we have a discrete $Z_5$ symmetry instead of the $Z_2$ used previously. This trivial modification turns out to have dramatic implications which are important enough to
be worth pointing out.
To begin with, it requires two Higgs doublets $H_u$ and $H_d$, which together with the leptoquark $S_3$, are charged non-trivially under $Z_5$. The explanation for $R_{K^{(*)}}$ is again connected to the origin of the Yukawa couplings and to CKM entries and mass ratios of SM fermions. An important qualitative difference between the present $Z_5$ model arises in the different topology of the diagrams that generate the effective couplings of $S_3$ to the SM fermions, appearing suppressed by two powers of vector-like fermion masses and also leading to $S_3$ not coupling to electrons. These differences play a crucial role in phenomenology, both pushing the mass of the leptoquark down, making it observable at the LHC,
while also eliminating the Lepton Flavour Violating (LFV) bound on $\mu \to e \gamma$ as the leading constraint of the present model, solving a main issue of the $Z_2$ implementation in \cite{deMedeirosVarzielas:2018bcy}.

As the present model can only consistently account for $R_{K^{(*)}}$ with light leptoquark masses and relatively large couplings, the bound from $B_s - \bar{B}_s$ mixing becomes relevant. The LFV bound on $\tau \to \mu \gamma$ also restricts the parameters of the model around the same order of magnitude as needed for $R_{K^{(*)}}$. This makes the model extremely predictive, being testable from updates to collider searches, $B_s - \bar{B}_s$ mixing or $\tau \to \mu \gamma$ in the near future.

Leptoquark extensions of the SM are motivated regardless of $R_{K^{(*)}}$, and controlling the couplings of the leptoquarks to the SM is important (due to predictivity and also proton decay). The present model is a noteworthy example, as in the phenomenologically interesting limit of light leptoquark masses and relatively large couplings, it has all the leptoquark couplings given in terms of known quantities and three model parameters $c_{\mu\tau}$, $\lambda_0$ and $M$, which can be constrained experimentally.

The layout of the paper now follows.
The model and a convenient basis for the discussion concerning the Yukawa couplings is introduced in Section \ref{sec:model}.
Then we consider the leptoquark couplings in the mass basis in Section \ref{sec:leptoquark_couplings}.
In Section \ref{sec:pheno} we analyse the phenomenological consequences arising from the leptoquark.
In Section \ref{sec:conc} we present the conclusions.

\section{The model \label{sec:model}}

The field and symmetry content of the model is presented in Table~\ref{tab:funfields1}.
The SM fermions and singlet neutrinos are neutral under a $Z_5$ symmetry, whereas the remaining fields all carry a charge under $Z_5$.
We have a fourth vector-like family,
two Higgs scalar doublets $H_u$, $H_d$, a SM singlet scalar $\phi$ and a scalar leptoquark $S_3$ that is an anti-triplet of $SU(3)_c$ and a triplet of $SU(2)_L$.

\begin{figure}[ht]
\centering

\hspace*{1ex}
	\includegraphics[scale=0.2]{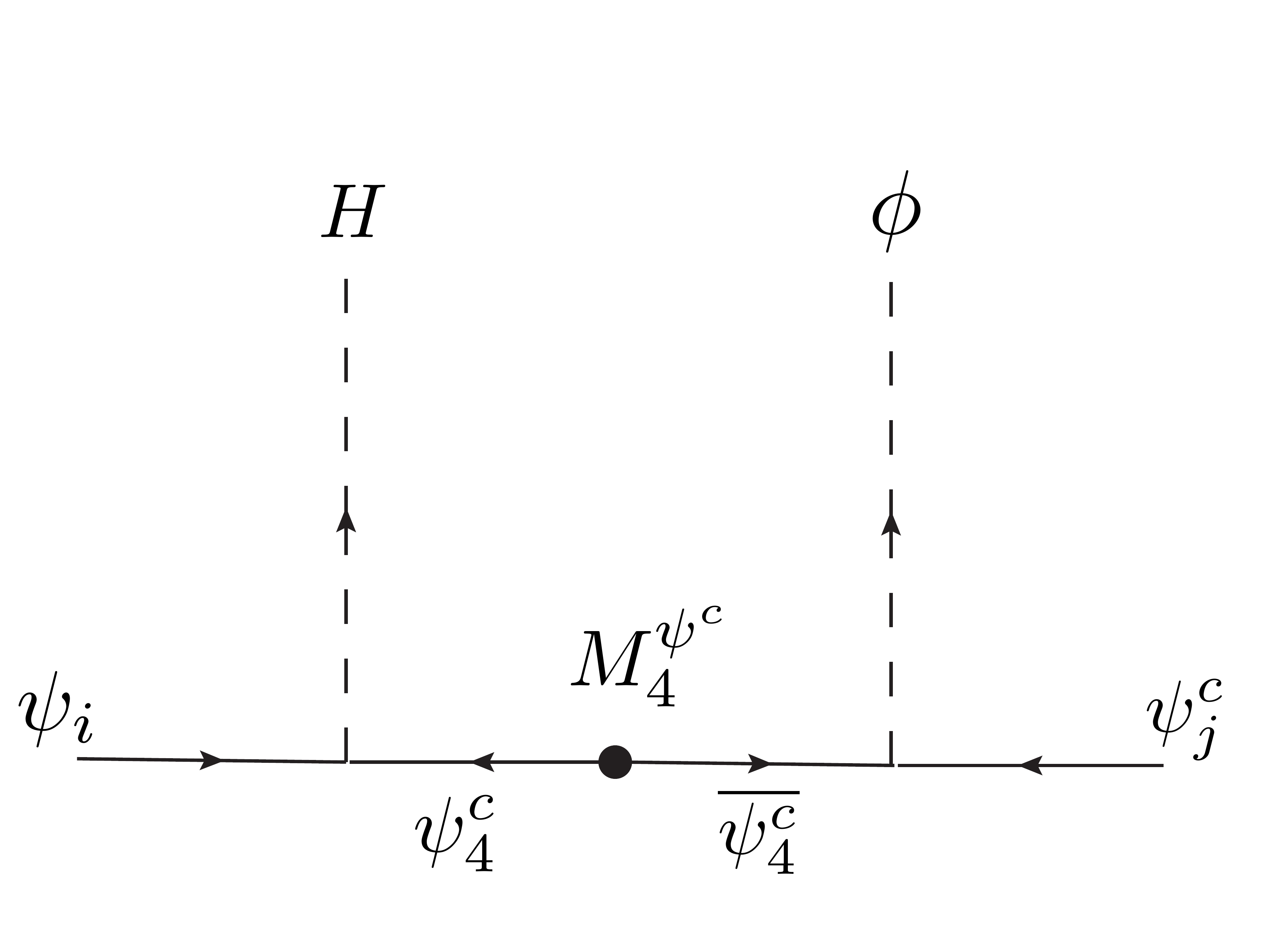}
	\includegraphics[scale=0.2]{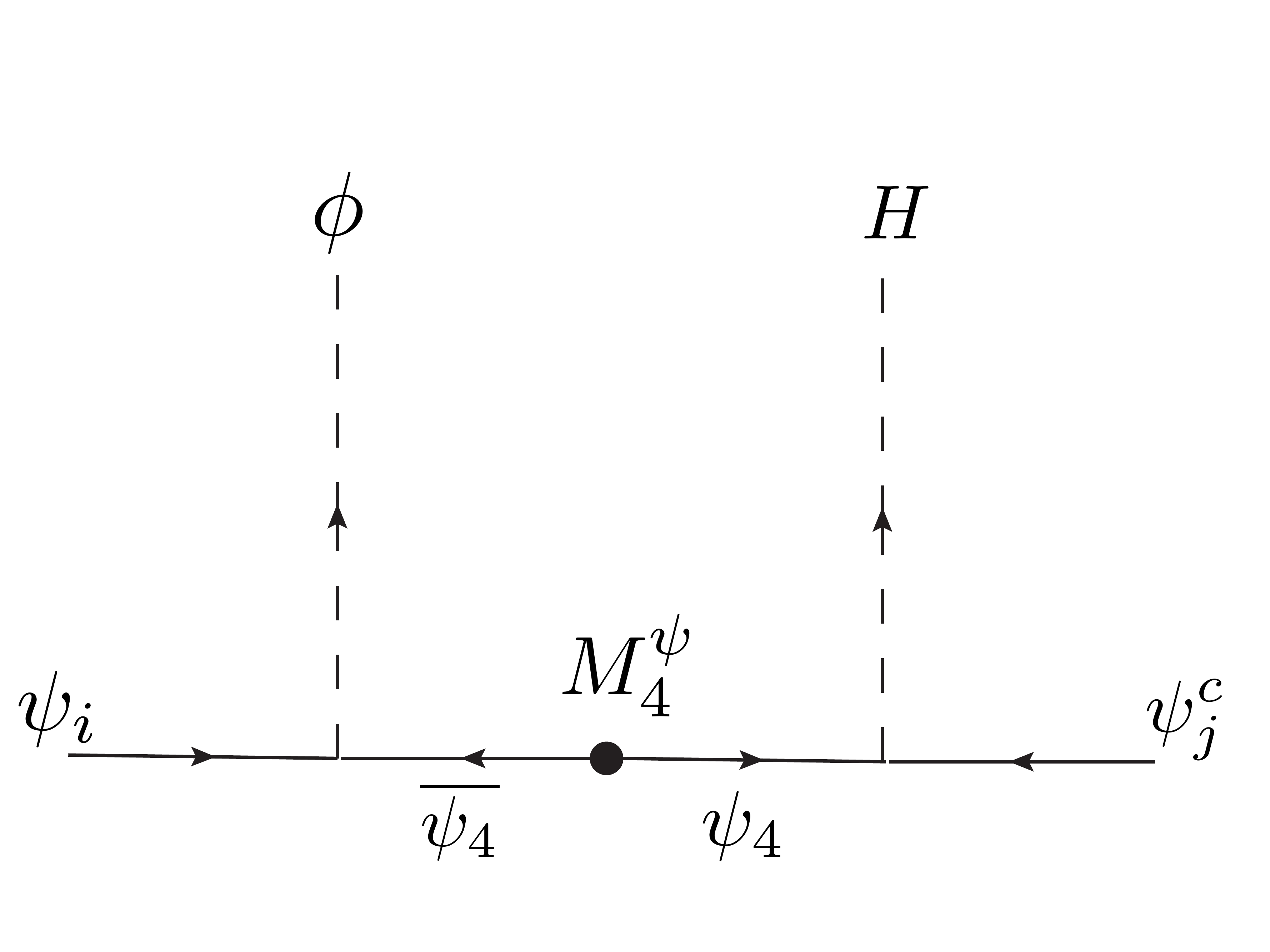}
\caption{Diagrams leading to effective Yukawa couplings between SM fermions and $H=H_u,H_d$.}
\label{Fig1}
\end{figure}

Since the SM chiral fermions are neutral under the $Z_5$, and the $H_u$ and $H_d$ are charged under $Z_5$,
renormalisable Yukawa couplings are forbidden. However Yukawa couplings involving the fourth family fermions
are allowed, plus other Yukawa couplings involving $\phi$. As a result of these couplings, 
effective Yukawa couplings involving the SM chiral arise from diagrams shown in Fig.~\ref{Fig1}. Note 
that two Higgs doublets $H_{u,d}$ are needed with identical $Z_5$ charge and opposite hypercharge, rather than the one Higgs doublet $H$ of the Standard Model (SM).

The leptoquark $S_3$ has coupling at renormalisable level to quarks and leptons, but only to the 4th family
\be
\lambda_4 S_3 Q_4 L_4 \,,
\label{lambda}
\ee
in left-handed Weyl notation. The dangerous di-quark couplings $S^\dagger Q Q$, $S^\dagger Q Q_4$ and  $S^{\dagger}_3 Q_4 Q_4$ are not allowed by $Z_5$, which alleviates issues that leptoquark models can have with too fast proton decay (see e.g. \cite{Assad:2017iib}).
The chiral fermions (neutral under $Z_5$) couple to the leptoquark, but only at the effective level. The effective couplings involve the  renormalisable leptoquark coupling of Eq.~\ref{lambda}, as seen in the diagram in Fig.~\ref{Fig2}.

\begin{table}
\centering
\begin{tabular}{| l c c c c |}
\hline
Field & $SU(3)_c$ & $SU(2)_L$ & $U(1)_Y$ &$Z_5$\\ 
\hline \hline
$Q_{i}$ 		 & ${\bf 3}$ & ${\bf 2}$ & $1/6$ & $0$ \\
$u^c_{i}$ 		 & ${\overline{\bf 3}}$ & ${\bf 1}$ & $-2/3$ & $0$\\
$d^c_{i}$ 		 & ${\overline{\bf 3}}$ & ${\bf 1}$ & $1/3$ & $0$\\
$L_{i}$ 		 & ${\bf 1}$ & ${\bf 2}$ & $-1/2$ & $0$\\
$e^c_{i}$ 		 & ${\bf 1}$ & ${\bf 1}$ & $1$ & $0$\\
$\nu^c_{i}$         & ${\bf 1}$ & ${\bf 1}$ & $0$ & $0$\\
\hline
\hline
$Q_{4}$ 		 & ${\bf 3}$ & ${\bf 2}$ & $1/6$ & $1$\\
$u^c_{4}$ 		 & ${\overline{\bf 3}}$ & ${\bf 1}$ & $-2/3$ & $1$\\
$d^c_{4}$ 		 & ${\overline{\bf 3}}$ & ${\bf 1}$ & $1/3$ & $1$\\
$L_{4}$ 		 & ${\bf 1}$ & ${\bf 2}$ & $-1/2$ & $1$\\
$e^c_{4}$ 		 & ${\bf 1}$ & ${\bf 1}$ & $1$ & $1$\\
$\nu^c_{4}$         & ${\bf 1}$ & ${\bf 1}$ & $0$ & $1$\\
\hline
\hline
$\overline{Q_{4}}$ 		 & $\overline{{\bf 3}}$ & $\overline{{\bf 2}}$ & $-1/6$ & $-1$\\
$\overline{u^c_{4}}$ 		 & ${{\bf 3}}$ & ${\bf 1}$ & $2/3$ & $-1$\\
$\overline{d^c_{4}}$ 		 & ${{\bf 3}}$ & ${\bf 1}$ & $-1/3$ & $-1$\\
$\overline{L_{4}}$ 		 & ${\bf 1}$ & $\overline{{\bf 2}}$ & $1/2$ & $-1$\\
$\overline{e^c_{4}}$ 		 & ${\bf 1}$ & ${\bf 1}$ & $-1$ & $-1$\\
$\overline{\nu^c_{4}}$         & ${\bf 1}$ & ${\bf 1}$ & $0$ & $-1$\\
\hline
\hline
$\phi$ & ${\bf 1}$ & ${\bf 1}$ & $0$ &$1$ \\
\hline
\hline
$S_3$ &  $\overline{\bf 3}$ & ${\bf 3}$ & $1/3$ & $3$\\
\hline
\hline
$H_u$ & ${\bf 1}$ & ${\bf 2}$ & $1/2$ &$-1$ \\
$H_d$ & ${\bf 1}$ & ${\bf 2}$ & $-1/2$ & $-1$\\
\hline
\end{tabular}
\caption{The field and symmetry content of the model. The SM fermions are denoted
$\psi_i=Q_i,L_i$ (left-handed), 
$\psi^c_i =u^c_i,d^c_i,e^c_i$ (right-handed),
and we add right-handed neutrinos $\nu^c_i$
($i=1,2,3$ for all chiral fermions, which are $Z_5$-neutral).
The fermion content is completed by the vector-like family charged under $Z_5$, $\psi_4, \psi^c_4$,
$\overline{\psi_4},\overline{\psi^c_4}$.
The scalars are all charged under $Z_5$, where we have
the $Z_5$-breaking field $\phi$,
two electroweak doublets $H_u$, $H_d$ and the leptoquark $S_3$.}
\label{tab:funfields1}
\end{table}

The respective renormalisable Lagrangian can be written as
\bea
{\cal L}^{ren} &=&
y^{\psi}_{i4}H  \psi_i {\psi^c_4} 
+  y^{\psi}_{4i}H {\psi_4} \psi^c_i+x^{\psi}_{i}\phi \psi_i \overline{\psi_4} 
+ x^{\psi^c}_{i}\phi \psi^c_i \overline{\psi^c_4}\nonumber \\
&+& M^{\psi}_{4}\psi_4 \overline{\psi_4}
+ M^{\psi^c}_{4}\psi^c_4 \overline{\psi^c_4}
+\lambda_4 S_3  Q_4 L_4 \,,
\label{Lag_ren}
\eea
where in rather compact notation, $H$ denotes $H_u, H_d$ and $\Psi$ stands for charged leptons as well as up and down quarks.

We start by considering the quarks, where we consider the convenient basis for the $Q_i, d^c_i, u^c_i$ ($i=1,\ldots 3$) where 
$x^{Q}_{1,2}=0$, $y^{u}_{41,42}=0$, $y^{d}_{41,42}=0$ (using the same notation as in \cite{King:2018fcg}).
In this basis we can still rotate the two lighter to get $x^{u^c}_1=0$, $x^{d^c}_1=0$ and $y^{u}_{14}=0$
We can not set $y^{d}_{14}= 0$ without loss of generality, as the rotations of the $Q_i$ are already exhausted.
Then finally the matrices of quark Yukawa couplings are given by
\be
	 \pmatr{
	&u^c_1& u^c_2&u^c_3&u^c_4&\overline{Q_4}\\ 
	\hline
	Q_1|&0&0&0&0 &0\\
	Q_2|&0&0&0&y^{u}_{24}H^u &0\\
	Q_3|&0&0&0&y^{u}_{34}H^u &x^{Q}_{3}\phi\\ 
	Q_4|& 0 & 0 & y^{u}_{43}H^u&0&M^{Q}_{4}\\ 
	\overline{u^c_4}|&0&x^{u^c}_{2}\phi&x^{u^c}_{3}\phi& M^{u^c}_{4}&0},
	 \pmatr{
	&d^c_1& d^c_2&d^c_3&d^c_4&\overline{Q_4}\\ 
	\hline
	Q_1|&0&0&0&y^{d}_{14}H^d &0\\
	Q_2|&0&0&0&y^{d}_{24}H^d &0\\
	Q_3|&0&0&0&y^{d}_{34}H^d &x^{Q}_{3}\phi\\ 
	Q_4|& 0 & 0 & y^{d}_{43}H^d&0&M^{Q}_{4}\\ 
	\overline{d^c_4}|&0&x^{d^c}_{2}\phi&x^{d^c}_{3}\phi& M^{d^c}_{4}&0} \,.
	\label{M^quark_an_basis}
\ee

In the basis and notation of \cite{King:2018fcg}, the mass insertion diagrams in Fig.~\ref{Fig1} lead, for $\langle \phi \rangle \ll M_4^Q$, to the quark Yukawa matrices
\bea
y^{u}_{ij}&=&  \pmatr{0 & 0 & 0 \\
0 & y^{u}_{24} x^{u^c}_{2} & y^{u}_{24} x^{u^c}_{3}\\
0 & y^{u}_{34} x^{u^c}_{2} & y^{u}_{34} x^{u^c}_{3}
}
\frac{\langle \phi \rangle }{M^{u^c}_{4}}
+\pmatr{
0 & 0 & 0 \\
0 & 0 & 0 \\
0 & 0 & x^{Q}_{3} y^{u}_{43}} 
\frac{\langle \phi \rangle }{M^{Q}_{4}} \nonumber \,, \\
y^{d}_{ij}&=&  \pmatr{
0 & y^{d}_{14} x^{d^c}_{2} & y^{d}_{14} x^{d^c}_{3} \\
0 & y^{d}_{24} x^{d^c}_{2} & y^{d}_{24} x^{d^c}_{3}\\
0 & y^{d}_{34} x^{d^c}_{2} & y^{d}_{34} x^{d^c}_{3}
}
\frac{\langle \phi \rangle }{M^{d^c}_{4}}
+\pmatr{
0 & 0 & 0 \\
0 & 0 & 0 \\
0 & 0 & x^{Q}_{3} y^{d}_{43}} 
\frac{\langle \phi \rangle }{M^{Q}_{4}} \,.
\label{Yuk_mass_insertion_ud}
\eea

We note that the effective Yukawa matrices are the sum of two matrices with a first column of zeros, adding up to rank 2 matrices, so the first family will have zero mass.
If one of the two Yukawa terms in each of the expressions in Eq.~\ref{Yuk_mass_insertion_ud} 
are dropped this would lead to rank 1 matrices, with the second family becoming massless as well. 
This observation 
suggests a natural explanation of the hierarchical smallness of the lighter family masses compared to the masses of the heaviest family, namely that 
one term dominates over the other one. 
This was called ``messenger dominance'' in  \cite{Ferretti:2006df}. 
To account for small $V_{cb}$ in the quark sector, it is natural to assume that 
the left-handed quark messengers dominate over the right-handed messengers, $M_4^Q\ll M_4^{d^c},M_4^{u^c}$,
which was called ``left-handed messenger dominance'' in  \cite{Ferretti:2006df},
with the further assumption $M_4^Q\ll M_4^{d^c}\ll M_4^{u^c}$ reproducing the stronger mass hierarchy
in the up sector (as compared to the down sector). Assuming all this leads to $|V_{cb}|\sim m_s/m_b$ 
with $V_{ub}$, though naturally small, being unconstrained  \cite{Ferretti:2006df}.
However to explain the smallness of the Cabibbo angle requires further model building 
such as an $SU(2)_R$ symmetry \cite{Ferretti:2006df}, although here we assume its smallness is accidental.

Similarly we consider the charged leptons, which are also hierarchical. In a convenient basis of $L_i, e^c_i$ ($i=1,\ldots 3$),
we have $x^{L}_{1,2}=0$, $y^{e}_{41,42}=0$, and $x^{e^c}_1=0$, and $y^{e}_{14}=0$:
\be
	 \pmatr{
	&e^c_1& e^c_2&e^c_3&e^c_4&\overline{L_4}\\ 
	\hline
	L_1|&0&0&0&0 &0\\
	L_2|&0&0&0&y^{e}_{24}H^d &0\\
	L_3|&0&0&0&y^{e}_{34}H^d &x^{L}_{3}\phi\\ 
	L_4|& 0 & 0 & y^{e}_{43}H^d&0&M^{L}_{4}\\ 
	\overline{e^c_4}|&0&x^{e^c}_{2}\phi&x^{e^c}_{3}\phi& M^{e^c}_{4}&0},
	\label{M^lepton_an_basis}
\ee
leading in the mass insertion approximation, for $\langle \phi \rangle \ll M_4^L$ to 
\be
y^{e}_{ij}=  \pmatr{0 & 0 & 0 \\
0 & y^{e}_{24} x^{e^c}_{2} & y^{e}_{24} x^{e^c}_{3}\\
0 & y^{e}_{34} x^{e^c}_{2} & y^{e}_{34} x^{e^c}_{3}
}
\frac{\langle \phi \rangle }{M^{e^c}_{4}}
+\pmatr{
0 & 0 & 0 \\
0 & 0 & 0 \\
0 & 0 & x^{L}_{3} y^{e}_{43}} 
\frac{\langle \phi \rangle }{M^{L}_{4}} \,.
\label{Yuk_mass_insertion_e}
\ee

For the neutrinos, due to $\nu_i^c$ being neutral under SM and $Z_5$, large Majorana mass terms are allowed and lead to the seesaw mechanism. As the charged lepton Yukawa couplings are approximately diagonal in this basis, the PMNS comes mostly from neutrino contributions that arise after the seesaw. Nevertheless, the small rotation in the charged lepton sector controls the admixture of $\mu$ and $\tau$ contained in $L_3$.

The requirements $\langle \phi \rangle \ll M_4^{Q,L}$ should be relaxed due to the large couplings required to the top quark in particular. The more rigorous diagonalisation procedure leads to entirely comparable structures, essentially replacing the $x$ parameters with mixing angles \cite{King:2018fcg, deMedeirosVarzielas:2018bcy}. Of these angles, the two that are relevant to mention here are 
\bea
\sin \theta_{34}^Q &= s^{Q}_{34} &= \frac{x^Q_3 \langle \phi \rangle }{ \sqrt{(x^Q_3 \langle \phi \rangle )^2+(M^{Q}_{4})^2 }} \,, \label{rigorous_angle}\\
\sin \theta_{34}^L &= s^{L}_{34} &= \frac{x^L_3 \langle \phi \rangle }{ \sqrt{(x^L_3 \langle \phi \rangle )^2+(M^{L}_{4})^2 }} \,.
\eea

In compact form, the $3\times3$ Yukawa matrices are
\be
y^{u,e}_{ij}=  \pmatr{0 & 0 & 0 \\
0 & y^{u,e}_{22} & y^{u,e}_{23}\\
0 & y^{u,e}_{32} & y^{u,e}_{33}
}
\,,
\label{Yuk_effective_ue}
\ee
\be
y^{d}_{ij}=  \pmatr{0 & y^{d}_{12} & y^{d}_{13} \\
0 & y^{d}_{22} & y^{d}_{23}\\
0 & y^{d}_{32} & y^{d}_{33}
} \,.
\label{Yuk_effective_d}
\ee
We recall that $y_{ij}$ couplings with $i,j=1,2,3$ (but not $4$) are effective couplings, that can be expressed as functions of renormalisable couplings $x$ , $y_{i4}$, $y_{4j}$ and of the ratios between $\langle \phi \rangle$ and the respective mediator masses $M_4$, and this is particularly clear in the mass insertion approximation (Eqs.\ref{Yuk_mass_insertion_ud},\ref{Yuk_mass_insertion_e}).
With the assumption of left-handed messenger dominance, $y^{u,d,e}_{33}$ are larger than the other entries (with contributions with $M_4^{Q,L}$), leading larger masses for the third families of SM fermions and also enabling the use of the small angle approximation in diagonalising the matrices. We introduce then the angles $\theta_{ij}$ with $i,j=1,2,3$ (but not $4$) as the parameters involved in the diagonalising the matrices of Yukawa couplings with the effective couplings $y_{ij}$, in Eqs.\ref{Yuk_effective_ue}, \ref{Yuk_effective_d}
\be
\theta_{23}^{u,d,e} \simeq y^{u,d,e}_{23}/y^{u,d,e}_{33} \,,
\label{t23}
\ee
and similarly, for the down quarks we have
\be
\theta_{13}^{d} \simeq y^{d}_{13}/y^{d}_{33} \,,
\ee
\be
\theta_{12}^{d} \simeq y^{d}_{12}/y^{d}_{22} \,.
\ee
Within the approximation considered, the other mixing angles vanish (as do the first family masses). With the hierarchy between the masses of charm and top quarks (governed by $y^{u}_{22}/y^{u}_{33} \sim M^Q_4/M^{u^c}_4$) being stronger than the hierarchy of the masses of the strange and bottom quarks (governed by $y^{d}_{22}/y^{d}_{33} \sim  M^Q_4/M^{d^c}_4$), we establish $y^{d}_{23}/y^{d}_{33} \gg y^{u}_{23}/y^{u}_{33}$. The CKM mixing angles receive negligible contributions from the up sector, so we take as a good approximation, in the special basis we are working so far, that the Yukawa couplings of the up quarks are already diagonal  (approximately). This means that we are considering $Q_3$ to contain the top (the mass eigenstate, $t$) and its down-type counterpart (expressed in terms of down-type mass eigenstates through the CKM matrix).

As we don't consider the origin of neutrino masses, for the charged leptons we take the simple assumption that the respective Yukawa coupling matrix  is diagonalised by a small $\theta_{23}^e \sim m_\mu/m_\tau$. This is justified given Eq.~\ref{t23} and that $y^e_{33}$ comes from $\frac{\phi}{M^L_4}$ and $y^e_{23}$ from $\frac{\phi}{M^{e^c}_4}$ (see Eq.~\ref{Yuk_mass_insertion_e}). We write
\be
\theta_{23}^e \equiv c_{\mu \tau} m_\mu/m_\tau \,,
\label{t23_est}
\ee
defining $c_{\mu \tau}$ as a parameter which we expect to be $O(1)$, with the smallness of $\theta_{23}^e$ coming explicitly from $m_\mu/m_\tau$. In order for the model to remain consistent, this angle should remain small, e.g. $\theta_{23}^e \lesssim 0.3 \sim \pi/10$,  which corresponds to $c_{\mu \tau} \lesssim 5$.

This type of consideration leads to \cite{deMedeirosVarzielas:2018bcy}
\bea
\label{dmass}
u_1 &\approx  u_L\,, \quad d_1 =& V_{ud} d_L +  V_{us} s_L + V_{ub} b_L \,, \nonumber \\
u_2 &\approx  c_L\,, \quad d_2 =& V_{cd} d_L +  V_{cs} s_L + V_{cb} b_L \,, \nonumber \\
u_3 &\approx  t_L\,, \quad d_3 =& V_{td} d_L +  V_{ts} s_L + V_{tb} b_L \,,
\eea
and
\bea
\label{emass}
e_1&=& e_L \,, \nonumber \\
e_2&\simeq& (1 - (\theta_{23}^e)^2) \mu_L - \theta_{23}^e \tau_L \,, \nonumber \\
e_3&\simeq& \theta_{23}^e \mu_L + (1 - (\theta_{23}^e)^2) \tau_L \,.
\eea

We note that in the model, the PMNS angles are not predicted as we have not specified the neutrino sector, which would contribute to the PMNS angles.
The Cabibbo angle, which depends on $y^d_{14}/y^d_{24}$, is also not predicted in this construction.
Conversely, the other two CKM angles are predicted to be small in this model, through the smallness of $M_4^Q/M_4^{d^c}$.

Since the top quark Yukawa is large and demands
a large mixing with the $\phi$ fields, we expect an order unity effective quark mixing, but from the lepton side the respective factor is related to the $\tau$ Yukawa. In the presence of the SM Higgs sector we would therefore expect that if 
$\frac{\langle \phi \rangle}{M^{Q}_{4}} \sim 1$, then $\frac{\langle \phi \rangle }{M^{L}_{4}} \sim m_\tau/m_t$ \cite{deMedeirosVarzielas:2018bcy}, which would justify neglecting contributions suppressed by the larger $M^{L}_{4}$ when compared with contributions involving $M^{Q}_{4}$.
In the present model the Higgs sector includes $H_u$ and $H_d$.
Fig.~\ref{Fig1} illustrates the origin of the effective Yukawa couplings giving rise to the mass of the top quark and $\tau$ lepton. The respective renormalisable couplings appear in Eq.~(\ref{Lag_ren}), with $x$ couplings to the $\phi$ and $y$ couplings to $H_u$ and $H_d$. We then refer to Eq.~(\ref{Yuk_mass_insertion_ud}) to write for the top quark
\be
y_t \sim \frac{x^Q_3 y^u_{43} \langle \phi \rangle}{M^{Q}_{4}} \,,
\ee
and to Eq.~(\ref{Yuk_mass_insertion_e}) to write for the $\tau$ lepton
\be
y_\tau \sim \frac{x^L_3 y^e_{43} \langle \phi \rangle }{M^{L}_{4}} \,,
\ee
such that taking the ratio we obtain
\be
\frac{x^L_3 \langle \phi \rangle }{M^{L}_{4}} \sim \frac{x^Q_3 \langle \phi \rangle }{M^{Q}_{4}} \frac{y^u_{43} y_\tau}{y^e_{43} y_t} \,.
\ee
We consider in addition the usual $\tan \beta \equiv v_u/v_d$ parametrisation of the ratio between the VEVs of $H_u$ and $H_d$, in order to estimate
\be
\frac{x^L_3 \langle \phi \rangle }{M^{L}_{4}} \sim \frac{x^Q_3 \langle \phi \rangle }{M^{Q}_{4}} \frac{y^u_{43} m_\tau}{y^e_{43} m_t} \tan \beta \,.
\label{MQMLratio}
\ee
This serves to show that depending on the renormalisable couplings and $\tan \beta$, it is quite possible to have $\frac{\langle \phi \rangle}{M^{L}_{4}} \sim 1$. These considerations will become important when we come to consider the 
effective leptoquark couplings
in the Section \ref{sec:leptoquark_couplings}.

\section{Effective leptoquark couplings \label{sec:leptoquark_couplings}}

\begin{figure}[ht]
\centering
	\includegraphics[scale=0.25]{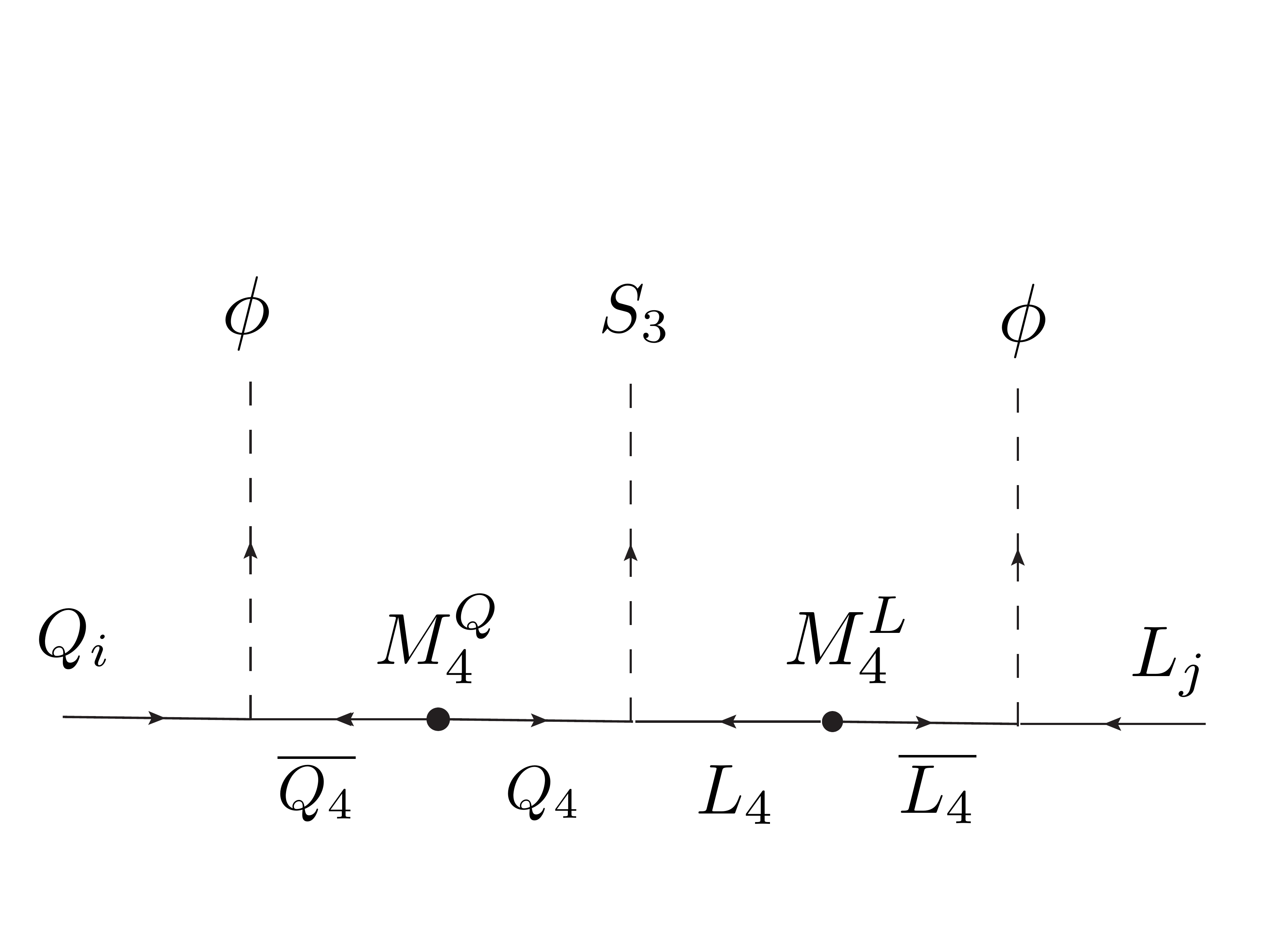}
\caption{Diagram in the model which leads to the effective leptoquark $S_3$ couplings in the mass insertion approximation.}
\label{Fig2}
\end{figure}

In the special basis where the lighter family is decoupled from the vector-like fermions (as seen in Eq.~(\ref{M^quark_an_basis},\ref{M^lepton_an_basis})), the effective leptoquark couplings from Fig.~\ref{Fig2} are
\be
\frac{x^{Q}_{3}\langle \phi \rangle }{M^{Q}_{4}}\frac{x^{L}_{3}\langle \phi \rangle }{M^{L}_{4}}
\lambda_4 S_3Q_3L_3 \ \equiv  \lambda^{ij}S_3Q_iL_j \,,
\label{leptoquark_couplings}
\ee
which is suppressed by two powers of vector-like fermion masses, another factor that justifies the detailed analysis leading to
Eq.~\ref{MQMLratio}.

The leptoquark only couples to a specific combination of quarks and a specific combination of leptons, which in this basis where
Eq.~\ref{leptoquark_couplings} holds, consists in just $Q_3$ and $L_3$, i.e.
\bea
\label{leptoquark_family_basis}
 \lambda^{ij} =
\frac{x^{Q}_{3}\langle \phi \rangle }{M^{Q}_{4}}\frac{x^{L}_{3}\langle \phi \rangle }{M^{L}_{4}}
\lambda_4
\left(
\begin{array}{ccc}
0 & 0 & 0 \\
0 & 0 & 0 \\
0 & 0 & 1
\end{array}
\right)
 \,.
\eea
Given that $Q_3$ and $L_3$ are the special quark and lepton flavour combinations coupling to the leptoquark $S_3$, we think of them as flavour eigenstates.
The effective leptoquark couplings to the SM quark and lepton mass eigenstates will arise from this single effective leptoquark coupling, by decomposing $Q_3$ and $L_3$ in terms of the respective mass eigenstates.

We want to express $Q_3$ and $L_3$ in terms of the mass eigenstates. In the leading order approximation where we considered up quark Yukawa couplings to be diagonal in the special basis, $Q_3$ contains $u_3 = t$ which coincides with the top quark (mass eigenstate) and the down-type combination within the $SU(2)_L$ doublet is obtained by the CKM matrix, namely $d_3 = V_{td} d +  V_{ts} s + V_{tb} b$ as shown in Eq.~\ref{dmass}.
$L_3$ contains an admixture of the $\tau$ and $\mu$ (mass eigenstates), but according to our assumptions, no $e$: $e_3 \simeq \theta_{23}^e \mu + (1 - (\theta_{23}^e)^2) \tau$ as shown in Eq.~\ref{emass}.

We are interested in the down quarks and charged lepton parts, in order to rewrite the couplings to $S_3$ as:
\be
\label{leptoquark_massbasis}
\lambda_{dl} =
\left(
\begin{array}{ccc}
\lambda_{de} & \lambda_{d\mu} & \lambda_{d\tau}  \\
\lambda_{se}  & \lambda_{s\mu}  & \lambda_{s\tau}     \\
\lambda_{be}  & \lambda_{b\mu}   & \lambda_{b\tau}  
\end{array}
\right) \,.
\ee

We recall that under the mass insertion approximation, $m_t$, $m_b$ and $m_\tau$ are approximately given by $x_3^Q y^u_{43} \phi/M_4^Q$, $x_3^Q y^d_{43} \phi/M_4^Q$, $x_3^L y^e_{43} \phi/M_4^L$ but are, in the more rigorous approach replaced by $s_{34}^Q y^u_{43}$, $s_{34}^Q y^d_{43}$, $s_{34}^L y^e_{43}$ shown explicitly in Eq.~\ref{rigorous_angle}.

In turn, the different $\lambda_{de}$, ... , $\lambda_{b\tau}$
arise from expressing the $Q_3$ and $L_3$ flavour eigenstates
through the mass eigenstates composing them, as in Eqs.(\ref{dmass}), (\ref{emass}).
Given Eq.~\ref{leptoquark_family_basis} only has couplings to $Q_3$ and $L_3$, it is clear that the $e$ column must vanish as $L_3$ does not contain $e$. Indeed, dropping the factors of $(\theta_{23}^e)^2 \sim (m_\mu/m_\tau)^2$ from the third column we have:
\be
\label{leptoquark_model}
\lambda_{dl} =
\lambda_0
\left(
\begin{array}{ccc}
0  &  \theta_{23}^e V_{td} & V_{td} \\
0  &  \theta_{23}^e V_{ts} & V_{ts} \\
0  &  \theta_{23}^e V_{tb} & V_{tb}
\end{array}
\right) \,,
\ee
where, from Eq.~\ref{leptoquark_couplings}, 
\be
\lambda_0 = \frac{x^{Q}_{3} \langle \phi \rangle}{M^{Q}_{4}}\frac{x^{L}_{3}\langle \phi \rangle }{M^{L}_{4}} \lambda_4 \equiv r \lambda_4 \,.
\label{rdef}
\ee
The effective coupling $\lambda_0$ is suppressed with respect to the renormalisable coupling $\lambda_4$.
Considering Eq.~\ref{rigorous_angle}, we place an upper bound of  $1/\sqrt{2}$ for $\frac{x^{Q}_{3}\langle \phi \rangle }{M^{Q}_{4}}$, and the same applies for the analogous leptonic $\frac{x^{L}_{3}\langle \phi \rangle}{M^{L}_{4}}$. Then we have as an upper bound on the ratio $r < \frac{1}{2}$.
\be
\lambda_0 \lesssim \frac{\lambda_4}{2} \,.
\label{lambda0_est}
\ee
In Section \ref{sec:pheno} we will consider bounds on the effective coupling $\lambda_0$, keeping in mind that it originates from renormalisable coupling $\lambda_4$.
Making use of Eq.~\ref{MQMLratio}, we note that depending on $\frac{y^u_{43} m_\tau}{y^e_{43} m_t} \tan \beta$, the upper bound can be saturated, provided
\be
\frac{y^u_{43} m_\tau}{y^e_{43} m_t} \tan \beta \sim 1
\ee
which can be understood as accounting for the hierarchy between the masses of the top quark and $\tau$ lepton not through an hiearchy of the vector-like $M^{Q,L}_4$ masses, but through an hierarchy of VEVs ($\tan \beta$) and of the Yukawa couplings $y^{u,e}_{43}$.

\section{Phenomenology \label{sec:pheno}}

The couplings in this Section are expressed in the mass basis of Eq.~\ref{leptoquark_massbasis}.

\subsection{Colliders}

A detailed analysis of the phenomenology of $S_3$ at hadron colliders was presented in \cite{Dorsner:2016wpm,Diaz:2017lit, Dorsner:2017ufx, Dorsner:2018ynv, Hiller:2018wbv, Schmaltz:2018nls}. 

In our model, the scalar leptoquark $S_3$ does not couple to $e$ and the couplings to $\mu$ are suppressed. There are constraints on the mass of the leptoquark from gluon-initiated pair production, which are mostly independent of the strength of the leptoquark coupling to fermions \cite{Diaz:2017lit, Dorsner:2018ynv, Hiller:2018wbv, Schmaltz:2018nls}. In our model the leptoquark couples dominantly to $b$ quark and $\tau$ lepton. The next leading couplings are suppressed by around $10^{-2}$, as couplings to $s$ quark are suppressed by $V_{ts}$, and couplings to $\mu$ lepton are suppressed by $\theta^e_{23}$. We take then the bounds quoted in \cite{Diaz:2017lit} which apply to leptoquarks coupling only to a single quark and to a single lepton family, namely $M > 1.4~\text{TeV}$ for $b \mu$ and $M > 1.0~\text{TeV}$ for $b \tau$ (the difference in bounds due to the sensitivity of experimental searches to $\mu$ and $\tau$). Single production bounds become relevant and start excluding $M > 1.5~\text{TeV}$ for coupling $\lambda_{b\tau} > 3$ \cite{Schmaltz:2018nls}.

With these collider bounds in mind, we present our analysis in terms of $\left( \frac{M}{1~\text{TeV}} \right)^2$ for convenience, considering that this leads to a factor of 2.25 for more conservative values of $M = 1.5~\text{TeV}$, a benchmark value which we also consider in some detail.

\subsection{$R_{K^{(*)}}$}

\begin{figure}[ht]
\centering

\hspace*{1ex}
	\includegraphics[scale=0.6]{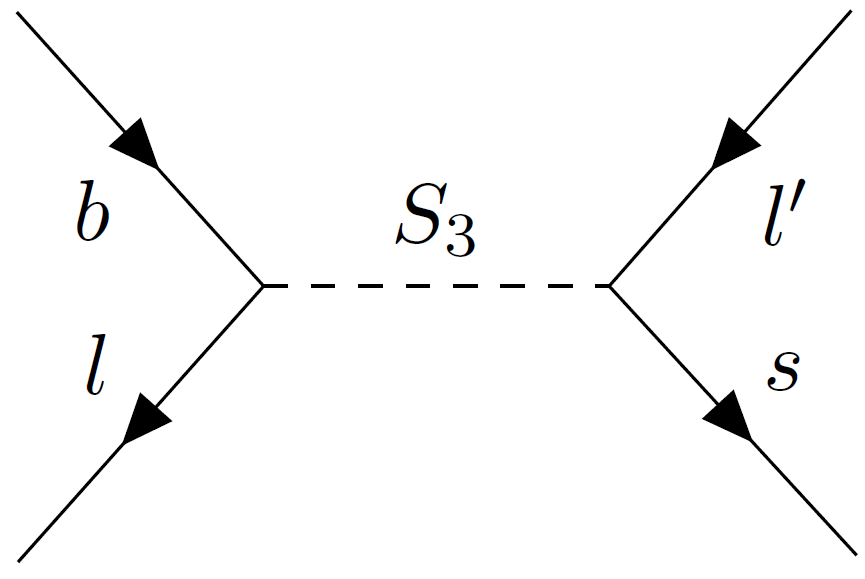}
\caption{Diagram of $S_3$ contributing to $R_{K^{(*)}}$ at tree level, with $l=l'=\mu$ in this model.}
\label{diag:RK}
\end{figure}

The leptoquark $S_3$ contributes to $R_{K^{(*)}}$ at tree level (see Fig. \ref{diag:RK}).
The requirement of getting $R_{K^{(*)}}$ from $S_3$ is \cite{Varzielas:2015iva, Hiller:2017bzc, Hiller:2018wbv}:
\be
\label{RKuirement}
\lambda_{b\mu} \lambda_{s\mu}^* - \lambda_{be}\lambda_{se}^* = \lambda_{b\mu} \lambda_{s\mu}^* \simeq 8.98 \times 10^{-4}  \left( \frac{M}{1~\text{TeV}} \right)^2 \,,
\ee
where the $S_3$ mass is $M$ and we take the electron couplings to be zero.

We can now insert the couplings from Eq.~\ref{leptoquark_model} to derive the impact of this requirement for our model
\be
(\theta^e_{23})^2 V_{ts} V_{tb}^* |\lambda_0|^2 
\simeq 8.98 \times 10^{-4}  \left( \frac{M}{1~\text{TeV}} \right)^2 \,,
\ee
and using Eq.~\ref{t23_est} we obtain
\be
c_{\mu \tau}^2 |\lambda_0|^2   
\simeq 6.35 \left( \frac{M}{1~\text{TeV}} \right)^2 \,.
\label{RKast}
\ee
We conclude from this that for our model to account for $R_{K^{(*)}}$ the leptoquark must be rather light,
otherwise the requirement forces large values of $c_{\mu\tau}$ or $\lambda_0$, which are expected to be small in our model (see Eq.~\ref{t23_est} and Eqs.~\ref{rdef},\ref{lambda0_est}).

\subsection{$B_s - \bar{B}_s$ mixing}

\begin{figure}[ht]
\centering

\hspace*{1ex}
	\includegraphics[scale=0.6]{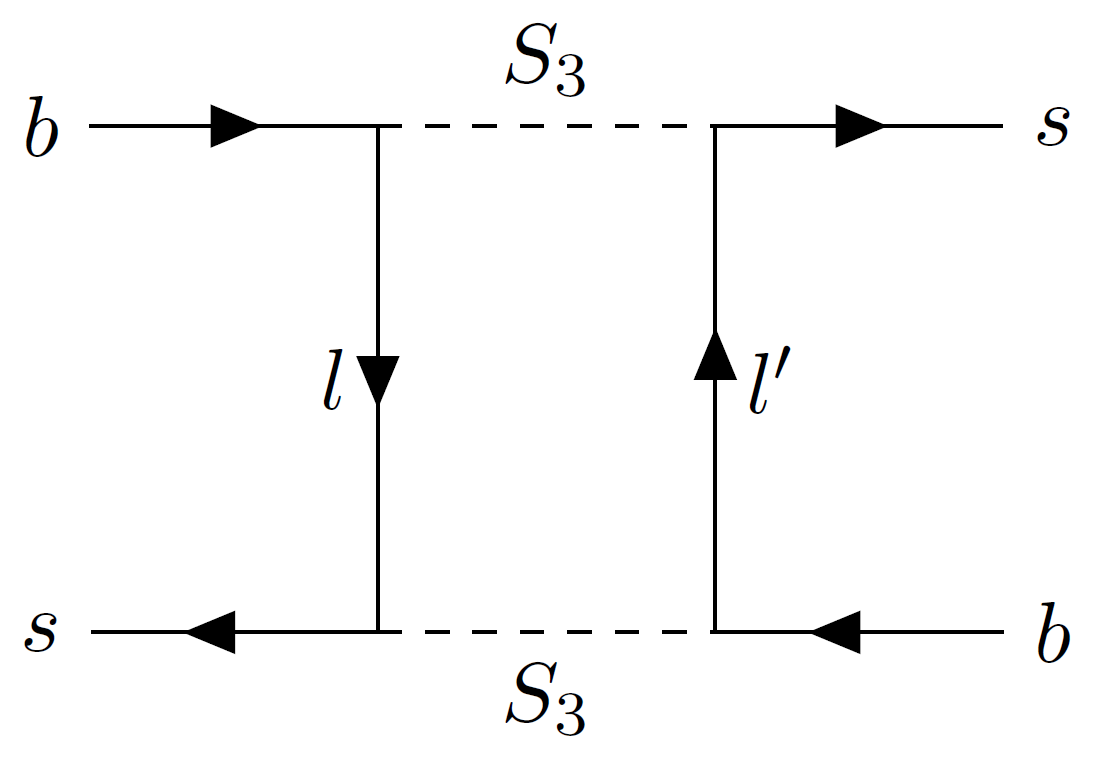}
\caption{Diagram of $S_3$ contributing to $B_s - \bar{B}_s$ mixing at 1-loop level.}
\label{diag:Bs}
\end{figure}

$S_3$ contributes to $B_s - \bar{B}_s$ mixing at 1-loop level (see box diagram in Fig. \ref{diag:Bs}).
The most strict constraint from $B_s - \bar{B}_s$ mixing on leptoquark couplings can be expressed as
\cite{Varzielas:2015iva}:
\be
(\lambda_{se}\lambda_{be}^*+\lambda_{s\mu}\lambda_{b\mu}^*+\lambda_{s\tau}\lambda_{b \tau}^*)^2 = (\lambda_{s\mu}\lambda_{b\mu}^*+\lambda_{s\tau}\lambda_{b \tau}^*)^2\lesssim 3.34 \times 10^{-3} \left( \frac{M}{1~\text{TeV}} \right)^2 \,,
\ee
where we take the electron couplings to be zero.
This depends on the fourth power of the coupling (instead of on the square of the coupling), which is due to the process ocurring at 1-loop level in a diagram with $S_3$ and leptons in the internal lines.
$B_s - \bar{B}_s$ mixing is sensitive to the $\tau$ couplings shown in Eq.~\ref{leptoquark_model}, which are not suppressed by $\theta_{23}^e$. We can rewrite
\be
(\lambda_{s\mu}\lambda_{b\mu}^*+\lambda_{s\tau}\lambda_{b \tau}^*)^2 =
\left( [(\theta_{23}^e)^2 + 1] V_{ts} V_{tb}^* \right)^2 |\lambda_0|^4  \simeq  1.60 \times 10^{-4} |\lambda_0|^4 \,,
\ee
assuming that we can drop $(\theta^e_{23})^2 \simeq 3.54 \times 10^{-3} c_{\mu\tau}^2$ 
(for $c_{\mu\tau} \lesssim 5$, $(\theta^e_{23})^2 \lesssim 0.09$),
 the number is coming from the suppression by $V_{ts}^2$. We therefore place a bound on $\lambda_0$ independently of $c_{\mu\tau}$
\be
|\lambda_0|^4 \lesssim 2.09 \left( \frac{M}{1~\text{TeV}} \right)^2 \,.
\ee

\begin{figure}[ht]
\centering

\hspace*{1ex}
	\includegraphics[scale=0.8]{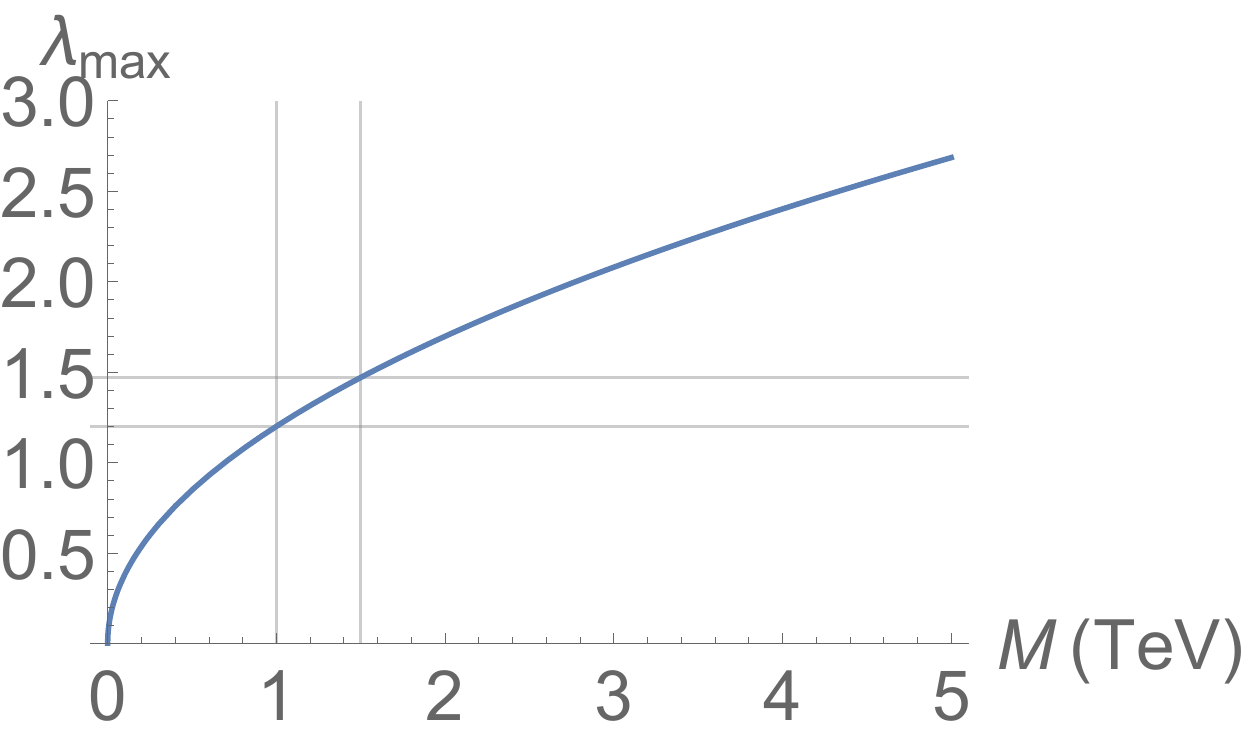}
\caption{The maximum value of $\lambda_{\text{max}}$ allowed for $\lambda_0$ due to the $B_s - \bar{B}_s$ bound. Gridlines show the values for $M=1~\text{TeV}$ and $M=1.5~\text{TeV}$.}
\label{Bs_M_lambda}
\end{figure}

The maximum value of $\lambda_0$ allowed for each $M$ is shown in Fig.~\ref{Bs_M_lambda}.

\subsection{LFV}

The leptoquark couplings are also constrained by lepton flavor violating (LFV) bounds \cite{Varzielas:2015iva}. Particularly stringent are $\mu \to e$ conversion processes such as the current bound on $\mathcal{B}(\mu \to e \gamma) = 5.7 \cdot 10^{-32}$ \cite{Adam:2013mnn}, leading to the constraint:
\begin{equation}
\label{LFV_mue}
| \lambda_{qe} \lambda_{q \mu}^* | \lesssim \frac{M^2}{(34~\text{TeV})^2} \,.
\end{equation}
Comparing Eq.~\ref{RKuirement} to Eq.~\ref{LFV_mue} indicates that there can be some tension if there is no hierarchy between $\mu$ and $e$ couplings. This was indeed an issue for the $Z_2$ model described in \cite{deMedeirosVarzielas:2018bcy}, where effective couplings originate from renormalisable couplings of each of the lepton families with the vector-like quark $Q_4$ and the leptoquark. In the present $Z_5$ model, this issue is resolved naturally as the only renormalisable coupling to the leptoquark is with $L_4$ and $Q_4$: the effective couplings to $e$ are entirely absent, as seen in Eq.~\ref{leptoquark_model}.
Indeed at leading order (with $m_e=0$) we have $\lambda_{qe} = 0$, automatically satisfying the $\mu \to e$ bounds.

For LFV bounds involving the $\tau$ lepton, $\mathcal{B}(\tau \to e \gamma) = 1.2 \cdot 10^{-7}$ \cite{Hayasaka:2007vc} and $\mathcal{B}(\tau \to \mu \gamma) = 4.4 \cdot 10^{-8}$ \cite{Aubert:2009ag} similarly constrain the respective leptoquark couplings:
\begin{equation}
\label{LFV_taue}
| \lambda_{qe} \lambda_{q \tau}^* | \lesssim \frac{M^2}{(0.6~\text{TeV})^2} \,,
\end{equation}
\begin{equation}
\label{LFV_taumu}
| \lambda_{q\mu} \lambda_{q \tau}^* | \lesssim 2.04 \left( \frac{M}{1~\text{TeV}} \right)^2 \,.
\end{equation}
With $\lambda_{qe} = 0$, the first of these bounds is also automatically satisfied. The latter bound
constrains the parameters of our model through the combination $c_{\mu\tau} |\lambda_0|^2$:
\be
| \lambda_{b\mu} \lambda_{b \tau}^* |
= \theta_{23}^e V_{tb}^2 |\lambda_0|^2 
\simeq c_{\mu\tau} \frac{m_\mu}{m_\tau} |\lambda_0|^2 \,,
\ee
therefore we have
\be
c_{\mu\tau} |\lambda_0|^2 \lesssim 34.3 \left( \frac{M}{1~\text{TeV}} \right)^2 \,.
\ee
When comparing this bound to Eq.~\ref{RKast} we understand that the present bound is automatically verified when the model explains $R_{K^{(*)}}$. Nevertheless, given the order of magnitude, future improvements to the experimental bound on
$\mathcal{B}(\tau \to \mu \gamma)$
will start constraining
$c_{\mu\tau}$, $\lambda_0$ and $M$.

\subsection{Discussion}

\begin{figure}[ht]
\centering

\hspace*{1ex}
	\includegraphics[scale=0.55]{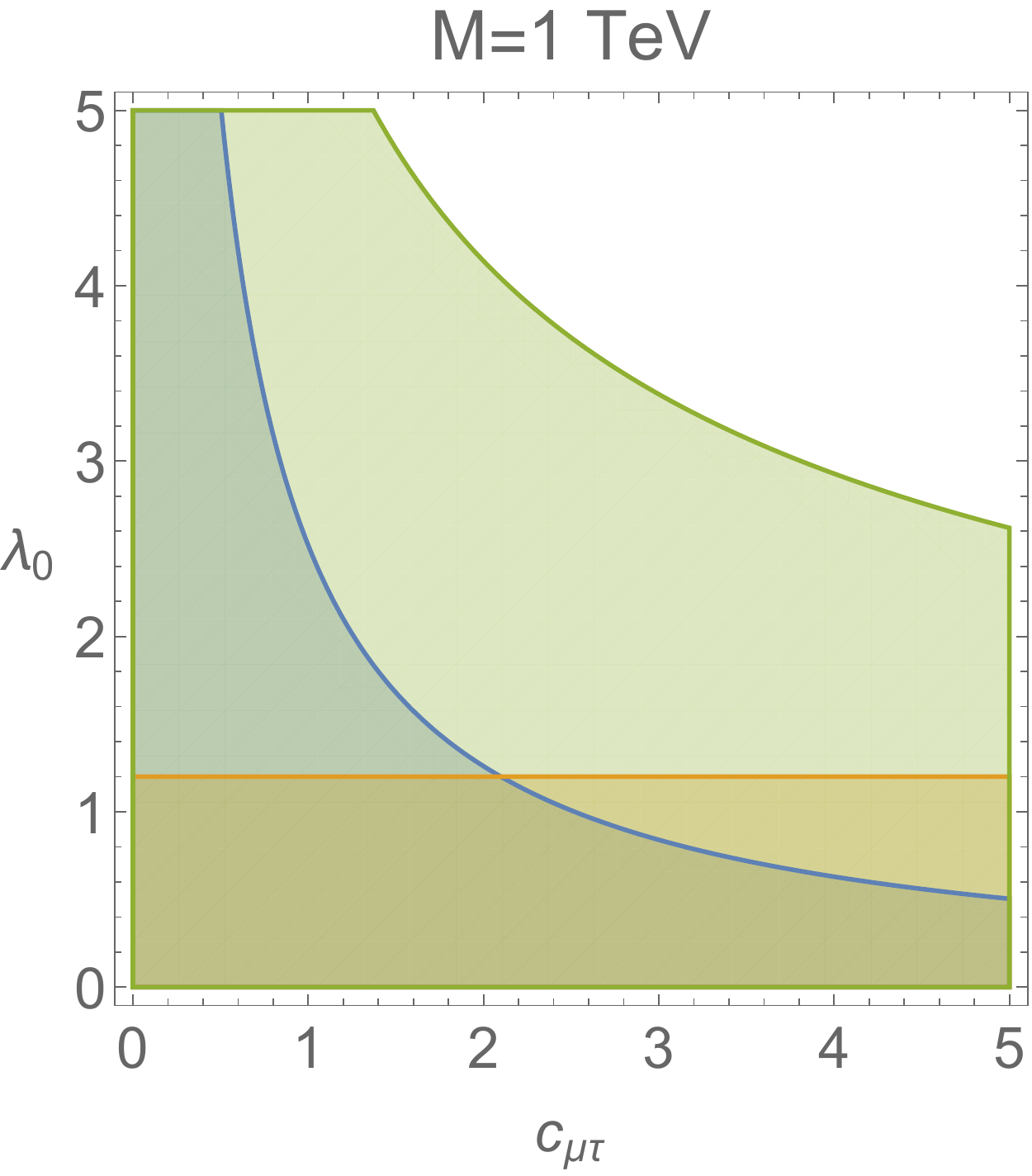}
\quad
	\includegraphics[scale=0.55]{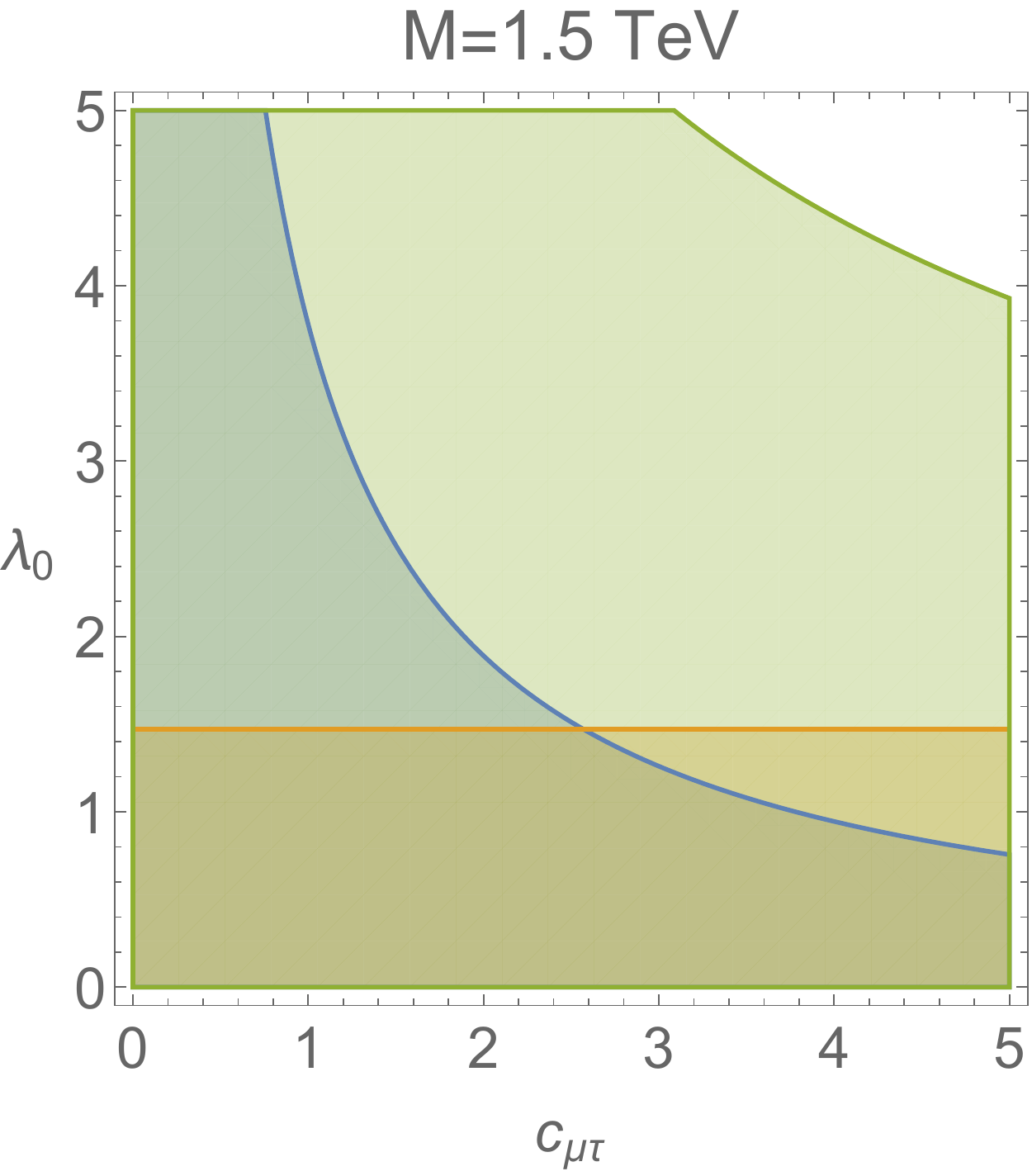}
\caption{The effect of $R_{K^{(*)}}$ (blue line), $B_s - \bar{B}_s$ (orange region) and $\tau \to \mu \gamma$ (green region) bounds on model parameters $c_{\mu\tau}$ and $\lambda_0$ respectively for $M=1~\text{TeV}$ (left panel) and for $M=1.5~\text{TeV}$ (right panel).}
\label{M_1}
\end{figure}

One can take Eq.~\ref{RKast} and recast it as an upper bound (this is the case if the leptoquark contribution to $R_{K^{(*)}}$ is lower than indicated), and compare this with the other bounds we obtained on the model parameters $c_{\mu \tau}$, $\lambda_0$ and $M$. We take for comparison the benchmark values of $M=1~\text{TeV}$ and $M=1.5~\text{TeV}$ and show the regions in Fig.~\ref{M_1} left and right panel respectively. 
Fig.~\ref{M_1} confirms that the present bound on $\tau \to \mu \gamma$ is not strong enough to constrain the model at the moment, but that it is expected to become relevant in the future.
In contrast, the $B_s - \bar{B}_s$ mixing bound is quite strong and excludes large values of $\lambda_0$, e.g. $\lambda_0 \lesssim 1.5$ for the values of $M$ considered. This in turn forces $c_{\mu \tau} \gtrsim 2$, as can be seen at the intersection in the left panel of Fig.~\ref{M_1}. From comparing the two panels in Fig.~\ref{M_1}, the minimum value of $c_{\mu\tau}$, depends on $M$. 

\begin{figure}[ht]
\centering

\hspace*{1ex}
	\includegraphics[scale=0.8]{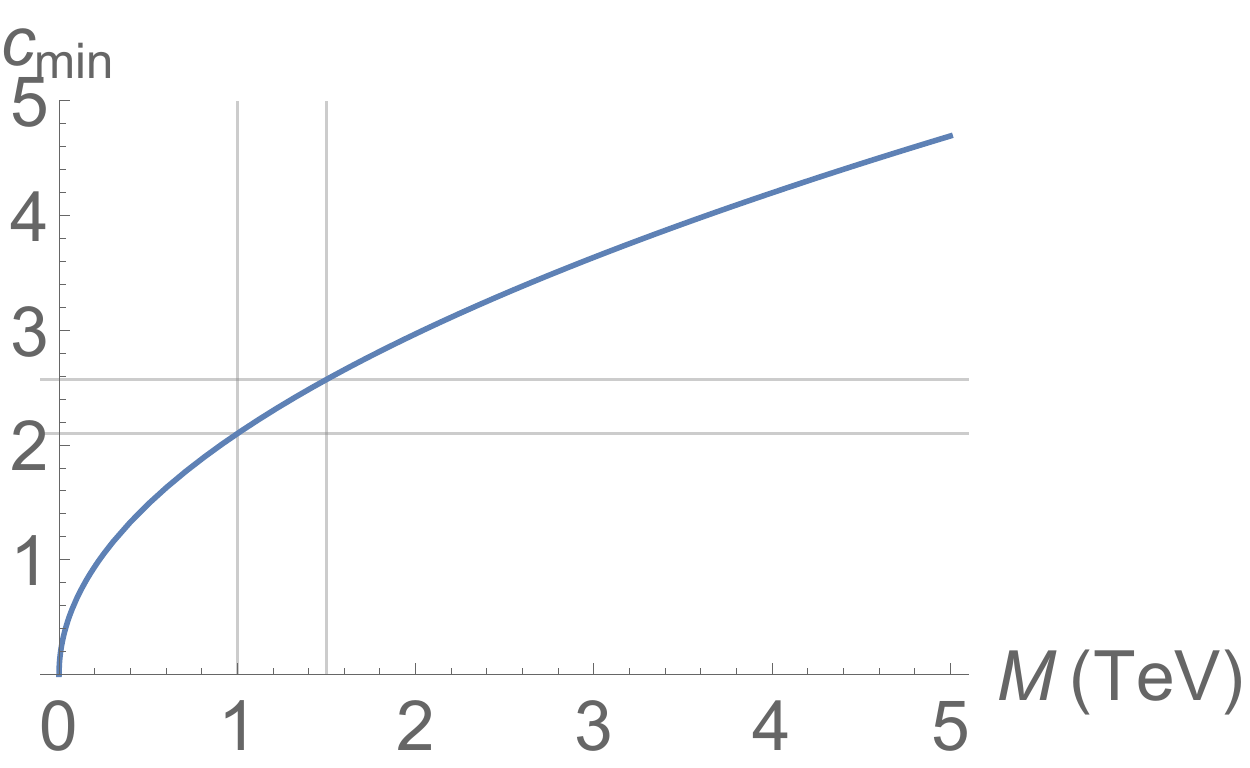}
\caption{The minimum value $c_{\text{min}}$ allowed for $c_{\mu\tau}$ allowed by simultaneously explaining $R_{K^{(*)}}$ and fulfilling the $B_s - \bar{B}_s$ bound. Gridlines show the values for $M=1~\text{TeV}$ and $M=1.5~\text{TeV}$.}
\label{M_c}
\end{figure}

In Fig.~\ref{M_c}, we plot the minimum value of $c_{\mu\tau}$, which we refer to as 
$c_{\text{min}}$, and show how it increases with $M$.
Large values of $c_{\mu \tau}$ are inconsistent with the underlying assumptions in the model (see Eq.~\ref{t23_est}), and Fig.~\ref{M_c} can be used to straightforwardly convert an upper bound on $c_{\mu\tau}$ into an upper bound on $M$,
e.g. $c_{\mu\tau} \lesssim 4.7$ leads to $M\lesssim 5~\text{TeV}$.

With further theoretical considerations, we can obtain even stricter upper bounds on $M$.
It is clear that in order to account for $R_{K^{(*)}}$, keeping $c_ {\mu\tau}$ fixed and decreasing $\lambda_0$ will also force the leptoquark to be lighter (see Eq.~\ref{RKast}). We recall that $\lambda_0 \equiv r \lambda_4$ (Eq.~\ref{rdef}), and conclude that keeping the renormalisable coupling $\lambda_4$ fixed, as $r$ decreases from its upper value $r=1/2$ the leptoquark becomes lighter. This theoretical consideration will constrain $M$ very strongly, as can be seen by taking e.g. $c_{\mu\tau} = 4.7$ and $r=1/2$, Eq.~\ref{RKast}, and taking the square root
\be
|\lambda_4|  
\simeq 1.07 \left( \frac{M}{1~\text{TeV}} \right) \,,
\ee
For given values of $\lambda_4 \sim O(1)$, $M$ takes similar values (for $M = 1$ to $1.5~\text{TeV}$ corresponds to $\lambda_4 = 1.07$ to $1.61$), and decreasing either $c_{\mu\tau}$ or $r$ forces $M$ to take lower values (in TeV) than $\lambda_4$.

We conclude that different theoretical considerations can be used to place robust upper bounds on $M$. An upper bound on $c_{\mu\tau} \lesssim 4.7$ would lead to $M \lesssim 5~\text{TeV}$ employing only the $B_s - \bar{B}_s$ mixing bound and the requirement of $R_{K^{(*)}}$. For the same values of $c_{\mu\tau}$, an even stronger bound of $M \lesssim 2~\text{TeV}$ can be placed if one considers additionally $r =0.1$ ($r \equiv \lambda_4/\lambda_0$).

\section{Conclusions \label{sec:conc}}

We have proposed a simple extension of the Standard Model in which the effective Yukawa couplings
and effective leptoquark couplings are related, leading to interesting constraints and predictions.
The model includes one scalar $SU(2)_L$ triplet leptoquark, two (rather than one) Higgs $SU(2)_L$ doublets, a scalar singlet, and vector-like fourth family of fermions (see Table~\ref{tab:funfields1}). All these are charged under a $Z_5$ symmetry under which the Standard Model chiral fermions are neutral, thereby preventing direct Yukawa couplings.
The $Z_5$ symmetry also forbids di-quark couplings to the leptoquark, alleviating the issue of proton decay.

The Yukawa couplings, forbidden by $Z_5$, appear effectively through the vacuum expectation value of $\phi$ breaking $Z_5$, from diagrams like Fig.~\ref{Fig1} revealing a single insertion of $\phi$. In contrast, the diagram in Fig.~\ref{Fig2} with double insertions of $\phi$, generates the
leptoquark couplings with the chiral fermions. Despite the distinct topologies, both types of couplings appear due to $\phi$ and the vector-like fermions.

The resulting model can account for the quark mixing angles and predicts the leading order leptoquark couplings to each down-type quark family in terms of the respective top quark CKM matrix element, $V_{td}$, $V_{ts}$ and $V_{tb}$. The relative strength of the coupling to $\mu$ is suppressed with respect to the coupling to $\tau$ through the mass ratio $m_\mu/m_\tau$. The leptoquark couplings thus follows the same hierarchy observed in charged lepton masses and in the quark mixing.

The model predicts lepton non-universality in $B$ to $K$ decays, depending on the leptoquark mass,
$V_{ts}$ and $m_\mu/m_\tau$. The model can only consistently explain the anomalies in $R_{K^{(*)}}$ for a
leptoquark mass close to the collider lower bound which we estimate to be about $1~\text{TeV}$.
There is no dedicated search for a leptoquark such as that predicted by our model 
with large couplings to $b$ and $\tau$, but also with suppressed couplings to $\mu$.
Constraints from $B_s - \bar{B}_s$ mixing become relevant
for such low leptoquark masses due to the large couplings to $\tau$, while $\mu \to e \gamma$ (and $\tau \to e \gamma$)
remain automatically under control due to the absence of leptoquark couplings to the electron in this model.
$\tau \to \mu \gamma$ in principle constrains the parameters of the model, but in practice less so than $B_s - \bar{B}_s$ mixing.

To summarise, the present model, which links the Yukawa couplings for the Higgs to those of the leptoquark,
is extremely predictive and can be tested in the near future. In order to consistently explain 
the current anomalies in $R_{K^{(*)}}$, the leptoquark mass needs to remain $O(1)~\text{TeV}$, putting it well within the reach of collider searches at the LHC. It can also be probed in $B_s - \bar{B}_s$ mixing (which significantly constrains the model) and eventually by improved bounds on $\tau \to \mu \gamma$.

\subsection*{Acknowledgements}

IdMV thanks Gudrun Hiller for helpful discussions and thanks the University of Southampton for hospitality during work on this project.

IdMV acknowledges funding from the Funda\c{c}\~{a}o para a Ci\^{e}ncia e a Tecnologia (FCT) through
the contract IF/00816/2015 and partial support by FCT through projects CFTP-FCT Unit 777 (UID/FIS/00777/2013), CERN/FIS-PAR/0004/2017 and PTDC/FIS-PAR/29436/2017 which are partially funded through POCTI (FEDER), COMPETE, QREN and EU.

S.\,F.\,K. acknowledges the STFC Consolidated Grant ST/L000296/1 and the European Union's Horizon 2020 Research and Innovation programme under Marie Sk\l{}odowska-Curie grant agreements Elusives ITN No.\ 674896 and InvisiblesPlus RISE No.\ 690575.

\end{document}